\documentclass[12pt]{article}

\usepackage{amssymb}

\usepackage{graphics}
\usepackage{epsfig}

\newcommand{\nn}{\nonumber}

\newcommand{\bq}{\begin{eqnarray} }
\newcommand{\eq}{\end{eqnarray} }

\begin{document}

\begin{titlepage}

\begin{flushright}
OSU-HEP-05-13\\
December 2005\\
\end{flushright}
\vspace*{0.5cm}
\begin{center}
{\Large {\bf{SU(2) Family Symmetry and the \\
Supersymmetric Spectrum}} }

\vspace*{1.5cm}
 {\large {\bf O.C. Anoka,$^{a,}$\footnote{E-mail
address:anoka@okstate.edu}
 K.S. Babu$^{a,}$\footnote{E-mail address:  babu@okstate.edu}
 and I. Gogoladze$^{a,b,}$\footnote{On a leave of absence from:
Andronikashvili Institute of Physics, GAS, 380077 Tbilisi, Georgia.
\\ \hspace*{0.5cm} E-mail address: ilia@physics.udel.edu}}}

 \vspace*{1.cm}
{\it $^a$ Department of Physics, Oklahoma State University\\
Stillwater, OK~74078, USA \\

$^b$ Department of Physics and Astronomy, University of Delaware\\
223 Sharp Laboratory, Newark, DE 19716, USA}
\end{center}

 \vspace*{1.5cm}

\begin{abstract}

We present a supersymmetric extension of the Standard Model with a
gauged $SU(2)$ family symmetry for the leptons.  It is shown that
this family symmetry can be consistently broken at the TeV scale
along with supersymmetry. If supersymmetry breaking is driven by
anomaly mediation, this model can provide positive squared masses
for the sleptons and thus cure the tachyon problem.  We analyze the
constraints and consequences of this scenario. A characteristic
feature of this model is the non--degeneracy of the first two family
sleptons. The model predicts large value of $\tan\beta$ and
observable $\tau \rightarrow e \gamma$ and $B \rightarrow \mu^+
\mu^-$ decay rates.
 \end{abstract}

\end{titlepage}

\newpage
\section{Introduction}

Family symmetries based on non--Abelian gauge groups may be useful in
addressing the mass hierarchy and flavor
mixings among quarks and leptons.  In a supersymmetric context, an
$SU(2)_F$ family symmetry, realized either as a global symmetry
\cite{dine,barbieri} or as a local symmetry \cite{bm1} can also provide
a solution to the SUSY flavor problem.  A concrete realization of
this idea would have the first two family fermions transforming as
doublets of $SU(2)_F$, while the third family fermions are singlets
of the group.  In the exact $SU(2)_F$ symmetric limit, the soft SUSY
breaking mass parameters for the first two families would  be
degenerate, alleviating the SUSY flavor problem significantly.  The
same setup can also explain the observed mass and mixing hierarchies
of quarks and leptons, including the neutrinos \cite{bm1}.

In this paper we propose a SUSY extension of the Standard Model (SM)
with a gauged family $SU(2)_F$ symmetry for the leptons.  When the
model is embedded into the anomaly mediated SUSY breaking (AMSB)
scenario \cite{Randall, Giudice}, we show that the model can
cure the tachyonic slepton problem.  This is our main
motivation.\footnote{Higher family symmetries, such as $SU(3)$ have
found application in solving the tachyon problem of AMSB \cite{abg}.
The realization presented here is somewhat simpler.} In order to
achieve this, we maintain asymptotic freedom of the $SU(2)_F$
interactions and insist that the symmetry breaks at the TeV scale
along with SUSY.  (If the symmetry were broken at a higher scale,
due to the ultraviolet insensitivity of AMSB scenario, there would be
no new contributions to the slepton masses.)  One of our
major tasks is to show that phenomenologically consistent $SU(2)_F$
gauge models broken at the TeV scale can be designed without
violating flavor changing neutral current (FCNC) constraints. The
main concern would be FCNC mediated by the $SU(2)_F$ gauge bosons
(we denote them as $V^a_\mu$). In our construction we show that owing to
approximate symmetries present in the model, excessive FCNC
processes do not occur.  We then show the consistency of the
symmetry breaking and study the salient features of the SUSY
spectrum.  Although the first two family sleptons are degenerate in
mass in the $SU(2)_F$ symmetric limit, symmetry breaking effects
lift this degeneracy by factors of order one.  While excessive
flavor violation is absent in the model, there are residual effects,
which we study.  It turns out that consistency of the model requires
$\tan\beta$ to be large, $\tan\beta \geq 40$. As a result, the model
predicts branching ratio for the process $B_s \rightarrow \mu^+
\mu^-$ very close to the current experimental limit.  While rare
decays involving the muon are suppressed in the model, decays such
as $\tau \rightarrow 3e$ and $\tau \rightarrow e \gamma$ are
allowed.  The branching ratio for the latter is found to be within
reach of future experiments.

A number of attempts to resolve the tachyonic slepton problem in
AMBS have appeared. For example, a non-decoupling universal bulk contribution to all the
scalar masses has been considered \cite{Pomarol, shadmi}. In this case the ultraviolet
(UV) sensitivity of the theory is no longer guaranteed.
Our TeV scale family symmetry resembles somewhat the phenomenology
of the universal bulk contributions, but differs from it in many crucial
aspects.  One common feature is that the neutral Wino is still the
lightest supersymmetric particle, which is nearly mass degenerate
with the lightest chargino.  The possibilities to detect such a
quasi-degenerate pair at the Tevarton Run 2 as well as at the LHC
was considered in Refs. \cite{Fengm}. The possibility that the neutral Wino
is the cold dark
matter candidate has also been studied \cite{cold}. Other approaches to
solving the problem include new TeV scale particle \cite{Chacko},
or interactions \cite{Allanach}, or $D$--term contributions from
a new $U(1)$ broken either at a high scale \cite{Jack, Arkani-Hamed}
or at the weak scale \cite{anoka}.
Non-decoupling effects of heavy fields at
higher orders have been studied \cite{Katz} in attempts to solve the tachyon problem.
In the model presented here the UV insensitivity of AMSB scenario is maintained.  There are a variety
of ways in which our model can be tested in forthcoming experiments.

The plan of this paper is as follows. In section \ref{sec22} we
present our model based on $SU(2)$ family symmetry. In section \ref{sec23} we analyze the Higgs
potential of the model. The SUSY spectrum is presented in section
\ref{sec24}. We discuss
our numerical results in section \ref{sec25} for a specific choice of
input parameters.  Here we show
explicitly the positivity of all slepton squared masses.
In section \ref{sec26}
we discuss the experimental implication of the model of immediate
interest. We give our conclusions in section \ref{sec27}.

\section{$SU(2)_F$ family symmetry\label{sec22}}

Our model, which is supersymmetric, is based on the gauge symmetry
$G\equiv SU(3)_{C}\times SU(2)_{L}\times U(1)_{Y}\times SU(2)_{F}$.
The matter fields of the model and their transformation properties
under the gauge symmetries are listed in Table \ref{SU2HTABLE1}.
Here $SU(2)_F$ is a horizontal symmetry that acts on the first two
families of lepton fields. The third family leptons transform as
singlets of $SU(2)_F$.  This symmetry is broken by the vacuum
expectation values of a pair of $SU(2)_F$ doublet Higgs fields
$\{\phi_u,~\phi_d$\} which are singlets of the SM gauge symmetry. In
the exact $SU(2)_F$ symmetric limit, the electron and the muon would
be degenerate. In order to lift this degeneracy, we introduce a pair
of $SU(2)_F$ singlet vector--like leptons \{$E$,
$E^c$\}.\footnote{Adding $SU(2)_F$ triplet Higgs fields that are
$SU(2)_L$ doublets is another option, but the $SU(2)_F$ beta
function will then be positive, which we wish to avoid.} The
$\Psi_N$ field is introduced for the cancelation of $SU(2)_F$ Witten
anomaly.  Note that the quark fields are all singlets of $SU(2)_F$.
This feature helps maintain asymptotic freedom of the $SU(2)_F$
interactions, a key ingredient for generating positive squared
masses for the sleptons when the model is embedded in
AMSB.\footnote{It is conceivable that there is a separate $SU(2)_F$
acting on the quark fields, which would make the model compatible
with quark--lepton symmetry.  We assume that such a family symmetry,
if present, is broken at a very high scale.}  Although we do not
explicitly use them, we note that if a pair of $SU(2)_F$ singlet
colored fields  $(3,1,-1/3) + (3^*, 1, 1/3)$ are added to the model,
along with the $\{E, E^c\}$ fields they would form complete ${\bf 5}
+ {\bf \overline{5}}$ of $SU(5)$ and would lead to unification of
the three gauge couplings associated with the SM gauge symmetries,
just as in the MSSM.

\begin{table}[h!]
\begin{center}
\begin{tabular}{|c|c|c|c|c|}\hline
\rule[4mm]{0mm}{0pt} Superfield &
 $ SU(3)_{C} $ & $SU(2)_{L}$& $U(1)_{Y}$ & $SU(2)_{F}$ \\

\hline \rule[4mm]{0mm}{0pt}$Q_{i}$&$3$&$2$&$\,\,\,\,\frac{1}{6}$&$1$\\

\hline
\rule[4mm]{0mm}{0pt}$u^c_i$&$\bar 3$&$1$&$-\frac{2}{3}$&$1$\\

\hline
\rule[4mm]{0mm}{0pt}$d^c_i$&$\bar3$&1&$\,\,\,\,\frac{1}{3}$&$1$\\

\hline \rule[4mm]{0mm}{0pt}$\psi=\pmatrix{L_e\cr
L_\mu}$&$1$&$2$&$-\frac{1}{2}$&$2$\\

\hline \rule[4mm]{0mm}{0pt}$\psi^c=\pmatrix{\mu^c\cr
-e^c}$&$1$&$1$&$+1$&$2$\\

\hline \rule[4mm]{0mm}{0pt}$L_\tau$&$1$&$2$&$-\frac{1}{2}$&$1$\\

\hline \rule[4mm]{0mm}{0pt}$\tau^{c}$&$1$&$1$&$+1$&$1$\\

\hline \rule[4mm]{0mm}{0pt}$H_{u}$&$1$&$2$&$\,\,\,\,\frac{1}{2}$&$1$\\

\hline \rule[4mm]{0mm}{0pt}$H_{d}$&1&$2$&$-\frac{1}{2}$&$1$\\

\hline\hline
\rule[4mm]{0mm}{0pt}$E$&$1$&$1$&$-1$&$1$\\

\hline \rule[4mm]{0mm}{0pt}$E^c$&$1$&$1$&$+1$&$1$\\
\hline
\rule[4mm]{0mm}{0pt}$\Psi_N=\pmatrix{N_1\cr N_2}$&$1$&$1$&$0$&$2$\\
\hline \rule[4mm]{0mm}{0pt}$\phi_{u}$&$1$&$1$&$\,\,\,\,0$&$2$\\
\hline \rule[4mm]{0mm}{0pt}$\phi_{d}$&1&$1$&$\,\,\,\,0$&$2$\\
\hline
\end{tabular}
\caption[Particle content and charge assignment of
the $SU(2)_F$ model.]%
{\footnotesize Particle content and charge assignment of the model.
The indices $i$ take values $i$ = $1-3$.}
  \label{SU2HTABLE1}
\end{center}
\end{table}

It should be noted that the particle content of any SUSY $SU(2)_F$
model that is asymptotically free is highly constrained. The
spectrum of Table \ref{SU2HTABLE1} is one of the few possibilities.
With the spectrum of Table \ref{SU2HTABLE1}, the $SU(2)_F$ gauge
beta function turns out to be negative, $\beta_{g_F}={-3g_F^3}/{(16\pi^2)}$.

We assume an unbroken $R$--parity as in the MSSM.  We take the
$R$--parity of the $\Psi_N$ field to be odd, similar to the SM
fermions. The superpotential of the model consistent with the gauge
symmetries reads as:
\begin{eqnarray}\label{SU23}
W&=&\left(Y_{u}\right)_{ij}Q_{i}H_{u}u_{j}^{c}
+\left(Y_{d}\right)_{ij}Q_{i}H_{d}d_{j}^{c}+f_{\mu}\psi^Ti\tau_2\psi^cH_{d}
+f_{\tau}L_\tau\tau^cH_d\nn\\&+&f_{\tau E}L_\tau
E^{c}H_d+f_{eE}E{\psi^c}^Ti\tau_2\phi_{d}+\mu
H_{u}H_{d}+\mu^{\prime} \phi_{u}^Ti\tau_2\phi_{d}+M_EEE^c.
\end{eqnarray}
Here we have suppressed the $SU(2)_L$ and $SU(3)_C$ indices, but
have shown the $SU(2)_F$ contraction explicitly.

The following additional terms are allowed in the superpotential by
the gauge symmetry:
\begin{equation}
W' = f_{eE}'E{\psi^c}^Ti\tau_2\phi_{u} + f_N \psi^T i \tau_2 \Psi_N
H_u.
\end{equation}
It turns out that the terms in $W'$ are not desirable.  The first
term in $W'$ would lead to unsuppressed flavor changing neutral
currents ($\mu \rightarrow 3 e$, $\mu \rightarrow e \gamma$ etc) via
the exchange of $SU(2)_F$ gauge boson, if $SU(2)_F$  is broken at
the TeV scale.  The second term would generate Dirac neutrino masses
for the SM neutrinos.  $\Psi_N$ cannot have a large Majorana mass,
as it is needed in the low energy theory for consistency.  We
observe that the terms in $W'$ can be eliminated consistently by
making use of a $Z_4$ symmetry present in $W$ and not in $W'$.
Under this $Z_4$ the fields with nontrivial transformation are:
\begin{equation}
\phi_u\rightarrow -i\phi_u,\,\,\,\phi_d\rightarrow
i\phi_d,\,\,\,E\rightarrow -iE,\,\,\,E^c\rightarrow
iE^c,\,\,\,L_{\tau}\rightarrow -iL_{\tau},\,\,\,\tau^c\rightarrow
i\tau^c,\,\,\,\Psi_N \rightarrow -\Psi_N.
\end{equation}
Although smallness of the couplings in Eq. (2) is adequate for
consistent phenomenology, we shall set $W'$ to zero by invoking this
$Z_4$ symmetry.

The neutrinos in the model get masses from the following
non-renormalizable operators (which we assume do not respect the
$Z_4$ symmetry)\footnote{If the $Z_4$ is broken in the $\nu_R$
sector, the breaking effects will show up only very weakly in the
effective low energy theory.}
\begin{eqnarray}
{L_\tau L_\tau H_u H_u \over M^2},\,\,\,\,\,\,\,\,
(\psi^Ti\tau_2\phi_{d,u}) L_{\tau}\frac{H_uH_u}{M^{\prime
2}},\,\,\,\,\,\,\,\,(\psi^Ti\tau_2\phi_{u,d})(\psi^Ti\tau_2\phi_{d,u})\frac{H_uH_u}{M^{\prime
\prime 3}}.\end{eqnarray} Consistent neutrino phenomenology can be
realized if the mass parameters in Eq. (4) are of order
$\{M,~M',~M''\} \sim \{10^{14},~10^9,~ 10^7\}$ GeV.  The fermionic
components of $\Psi_N$ fields will acquire masses through effective
operators of the form $(\Psi_N^T i \tau_2 \phi_{u,d})^2/M$, which are
expected to be of order sub--eV.

\section{Symmetry breaking\label{sec23}}
The symmetry breaking is assumed to follow the following
hierarchical pattern:
\begin{eqnarray}
SU(3)_C \times SU(2)_L \times U(1)_Y \times SU(2)_F\,\,\,   &~&
\stackrel{\langle  \phi_u \rangle, \, \langle  \phi_d \rangle
}{\longrightarrow}  \,\,
SU(3)_C \times SU(2)_L \times U(1)_Y \,\,\,\ \nonumber \\
&~& \stackrel{\langle  H_u \rangle, \, \langle  H_d \rangle
}{\longrightarrow}  \,\,\, SU(3)_C\times U(1)_{EM}. \nonumber
\end{eqnarray}
The tree level Higgs potential can be written as
 \begin{eqnarray}\label{SU26}
V(H_{u},H_{d},\phi_u,\phi_d)&=&(m_{H_{u}}^2+\mu^2)|H_{u}|^2+(m_{H_{d}}^2+\mu^2)|H_{d}|^2+B\mu(H_uH_d+
\rm{c.c.})\nn\\
&+&\frac{(g_2^2+g_1^2)}{8}(|H_{u}|^2-|H_{d}|^2)^2+\frac{g_2^2}{2}|H_uH_d|^2+\frac{{g_F}^2}{8}(|\phi_{u}|^2-|\phi_{d}|^2)^2\nn\\&+&\frac{{g_F}^2}{2}|\phi_u^Ti\tau_2\phi_d|^2
+(m_{\phi_{u}}^2+\mu^{\prime 2})|\phi_{u}|^2+
(m_{\phi_{d}}^2+\mu^{\prime
2})|\phi_{d}|^2\nn\\&+&B^{\prime}\mu^{\prime}(\phi_u^Ti\tau_2\phi_d+
\rm{c.c.}).
\end{eqnarray}
 The soft mass parameters $m_{H_u}^2$ and $m_{H_d}^2,m_{\phi_u}^2$ and
$m_{\phi_d}^2$
 are determined in terms of a single mass parameter
 $M_{aux}$ and the Yukawa and gauge couplings
 within AMSB.  We present these masses in the Appendix, see Eqs.
(63)-(66).
 The $B$ and $B^{\prime}$ parameters are assumed  to be a priori  free
 in the model.\footnote{In some special class of models, the $B$
parameters are
 determined in terms of $M_{aux}$ and the gamma functions of the Higgs fields.
 We do not assume these special values for $B$ and $B'$ here.}
 The Higgs fields acquire vacuum expectation values of the form
\begin{eqnarray}\label{SU27}
\langle H_u\rangle=\pmatrix{0\cr \upsilon_u},\,\,\,\,\langle
H_d\rangle=\pmatrix{\upsilon_d\cr0},\,\,\,\,\langle
\phi_u\rangle=\pmatrix{0\cr u_u},\,\,\,\,\langle
\phi_d\rangle=\pmatrix{u_d\cr 0}.
\end{eqnarray}
The desired symmetry breaking patterns can be achieved if the
hierarchy $u_u,\,u_d \gg \upsilon_u,\,\upsilon_d$ can be realized.

 Minimization of the Higgs potential $V$
 leads to the following conditions:
\begin{eqnarray}\label{SU28} \sin{2\beta}=
\frac{-2B\mu}{2\mu^2+m_{H_{u}}^2+m_{H_{d}}^2},\,\,\,\ \mu^2=
\frac{m_{H_{d}}^2-m_{H_{u}}^2\tan^2{\beta}}{\tan^2{\beta}-1}-\frac{M_{Z}^2}{2},
\end{eqnarray}
\begin{eqnarray}\label{SU29}
 \sin{2\beta^\prime}=
\frac{-2B^\prime\mu^\prime}{2\mu^{\prime
2}+m_{\phi_{u}}^2+m_{\phi_{d}}^2},\,\,\,\ \mu^{\prime 2}=
\frac{m_{\phi_{d}}^2-m_{\phi_{u}}^2\tan^2{\beta^\prime}}{\tan^2{\beta^{\prime}}-1}-
\frac{M_{V}^2}{2}.
\end{eqnarray}
Here we have introduced the notation $\tan\beta
=\upsilon_u/\upsilon_d$, $\tan{\beta^\prime}=u_u/u_d$, and
$u^2=u_u^2+u_d^2$. $M_{V}^2= \frac{{g_F}^2}{2}(u_u^2+u_d^2)$ is the
common mass of the three gauge boson associated with  $SU(2)_F$.

To find the physical Higgs boson masses, we parameterize the Higgs
fields (in the unitary gauge) as
\begin{eqnarray}\label{SU210}
&&H_u=\pmatrix{H^+\sin{\beta}\cr
\upsilon_u+\frac{1}{\sqrt{2}}(\phi_2+i\cos{\beta}\,\phi_3)},\,\,\,\,\langle
H_d\rangle=\pmatrix{\upsilon_d+\frac{1}{\sqrt{2}}(\phi_1+i\sin{\beta}\,\phi_3)\cr
H^-\cos{\beta}},\nonumber\\
&&\phi_u=\pmatrix{\phi^\prime\sin{\beta^\prime}\cr
u_u+\frac{1}{\sqrt{2}}(\phi_4+i\cos{\beta^\prime}\,\phi_5)},\,\,\,\,\phi_d=\pmatrix{u_d+\frac{1}{\sqrt{2}}(\phi_6+i\sin{\beta^\prime}\,\phi_5)\cr
\phi^{\prime \star}\cos{\beta^\prime}}.
\end{eqnarray}

The masses of the CP--odd Higgs bosons $\{\phi_3,\,\phi_5\}$ are
\begin{eqnarray}\label{SU2101}
m^2_A&=&\frac{-2B\mu}{\sin{2\beta}},\,\,\,\,\,
m^2_{A^{\prime}}=-\frac{2B^{\prime}\mu^{\prime}}{\sin{2\beta^\prime}}.
\end{eqnarray}

The mass matrices for the CP--even neutral Higgs bosons
$\{\phi_1,\,\phi_2\}$ and $\{\phi_4,\,\phi_6\}$ are decoupled. They
are given by
\begin{eqnarray}\label{SU211}
(\mathcal{M}^2)_{cp-even}&=&\pmatrix{m^2_A\cos^2{\beta}+M_Z^2\sin^2{\beta}&-\{m^2_A+M_Z^2\}\frac{\sin{2\beta}}{2}\cr
-\{m^2_A+M_Z^2\}\frac{\sin{2\beta}}{2}&m^2_A\sin^2{\beta}+M_Z^2\sin^2{\beta}},
\end{eqnarray}
\begin{eqnarray}\label{SU212}
(\mathcal{M}^{\prime
2})_{cp-even}&=&\pmatrix{m^{2}_{A^\prime}\cos^2{\beta^\prime}+M_V^2
\sin^2{\beta^\prime}&-\{m^{2}_{A^\prime}+M_V^2\}\frac{\sin{2\beta^\prime}}{2}\cr -\{m^{2}_{A^\prime}+
M_V^2\}\frac{\sin{2\beta^\prime}}{2}&m^{2}_{A^\prime}\sin^2{\beta^\prime}+M_V^2\sin^2{\beta^\prime}}.
\end{eqnarray}

Finally, the masses of the charged Higgs boson $H^{\pm}$ and that of
$\phi^{\prime}$ are given by
\begin{eqnarray}\label{SU213}
m^2_{H^{\pm}}=m_A^2+M_W^2,\,\,\,\,\,\,\,\,\,m^2_{\phi^{\prime}}=m_{A^\prime}^2+M_{V}^2,
\end{eqnarray}
The (complex) $\phi'$ fields are electrically neutral, but they are
``charged'' under $SU(2)_F$.

The Majorana mass matrix of the  neutralinos
$\{\tilde{B},\,\tilde{W_3},\,\tilde{H_d^0},\,\tilde{H_u^0},\,\tilde{B_{H}},\,\tilde{\phi_d^0},\,\tilde{\phi_u^0}\}$
is  found to be
\begin{eqnarray}\label{SU214}\small{
\mathcal{M}^{(0)}=\pmatrix{M_1&0&-\frac{\upsilon_d}{\sqrt{2}}g_1&\frac{\upsilon_u}{\sqrt{2}}g_1&0&0&0\cr
0&M_2&\frac{\upsilon_d}{\sqrt{2}}g_2&-\frac{\upsilon_u}{\sqrt{2}}g_2&0&0&0\cr
-\frac{\upsilon_d}{\sqrt{2}}g_1&\frac{\upsilon_d}{\sqrt{2}}g_2&0&-\mu&0&0&0\cr
\frac{\upsilon_u}{\sqrt{2}}g_1
&-\frac{\upsilon_u}{\sqrt{2}}g_2&-\mu&0&0&0&0\cr 0&0&0
&0&{\tilde{M}_2}&\frac{u_d}{\sqrt{2}}{g_F}&-\frac{u_u}{\sqrt{2}}{g_F}\cr
0&0&0&0&\frac{u_d}{\sqrt{2}}{g_F}&0&-\mu^{\prime}\cr
0&0&0&0&-\frac{u_u}{\sqrt{2}}{g_F}&-\mu^{\prime}&0}},
\end{eqnarray}
where $M_1,\,M_2\,$and$\,{\tilde{M}_2}$ are the gaugino masses for
$U(1)_Y,\,SU(2)_L \,$and$\,SU(2)_F$ which are listed for the case of
AMSB in Eq. (62) of Appendix \ref{appendix5}. The physical
neutralino masses $m_{\tilde{\chi}_i^0}$ ($i=$1--7) are obtained as
the eigenvalues of this mass matrix Eq. (\ref{SU214}).

In the basis $\{\tilde{W}^+,\,\tilde{H}_u^+\}$,
$\{\tilde{W}^-,\,\tilde{H}_d^-\}$, the chargino (Dirac) mass matrix
is
\begin{eqnarray}\label{SU215}
\mathcal{M}^{(c)}=\pmatrix{M_2&g_2\upsilon_d\cr
g_2\upsilon_{u}&\mu}.
\end{eqnarray}
Similarly, for the $SU(2)_F$ sector, we have
\begin{eqnarray}\label{SU2152}
\tilde{\mathcal{M}}^{(c)}=\pmatrix{{\tilde{M}_2}&{g_F}u_d\cr
{g_F}u_{u}&\mu^\prime}.
\end{eqnarray}
The three $SU(2)_F$ gauge boson masses are given by
\begin{eqnarray}\label{SU216}
\mathcal{M}_V^{2}=\frac{{g_F}^2}{2}(u_u^2+u_d^2).
\end{eqnarray}

\subsection{Lepton masses}
Now we describe how realistic lepton masses with $m_e \neq m_\mu$
are generated in the model. We have introduced $E$ and $E^c$ fields
in the superpotential Eq. (\ref{SU23}) for the purpose of breaking
$e-\mu$ degeneracy. These new fields mix with the usual leptons
leading to the mass matrix
\begin{eqnarray}\label{SU299}
\mathcal{L}_{mass}=\pmatrix{e&\mu&\tau&E}\pmatrix{f_{\mu}\upsilon_d&0&0&0\cr
0&f_\mu\upsilon_d&0&0\cr 0&0&f_\tau\upsilon_d&f_{\tau
E}\upsilon_d\cr f_{eE}u_d&0&0&M_E}\pmatrix {e^c\cr \mu^c\cr
\tau^c\cr E^c}.
\end{eqnarray}
The muon field  decouples from the rest of the leptons.  This
enables us to define an approximate muon number\footnote{since the
$Z_4$ symmetry is broken in the $\nu_R$ sector, muon number is only
an approximate symmetry in the model.} which guarantees that there
is no excessive FCNC processes involving the muon (for which the
experimental constraints are the most stringent). We are left with a
$3\times 3$ mass matrix for the $e$, $\tau$ and $E$ fields. The
eigenvalue equation can be solved using the hierarchy $m_e\ll
m_{\tau}\ll m_E$ and for $\frac{\upsilon_d}{M_E}\ll 1$
(corresponding to large $\tan\beta$) with the result
\begin{eqnarray}\label{ppp}
m_{\mu}&=&f_\mu\upsilon_d\nn\\
\frac{m_e}{m_\mu}&\simeq&\left[1+\frac{f_{eE}^2u_u^2}{M_E^2}\left(1+\frac{f_{\tau
E}^2}{f_\tau^2}\right)\right]^{-\frac{1}{2}}\nonumber\\
m_\tau &\simeq& f_\tau\upsilon_d\left[{1+\frac{f_{\tau E}^2f_{e
E}^2}{f_\tau^2}\frac{u_u^2}{M_E^2+f_{eE}^2u_u^2}}\right]^{\frac{1}{2}}\left[1+\Delta_\tau\right]\equiv
y_\tau\upsilon_d(1+\Delta_\tau),\nonumber\\
{m}_E &\simeq& \left[M_E^2+f_{eE}^2u_u^2\right]^{\frac{1}{2}}.
\end{eqnarray}
Note that $m_e\neq m_{\mu}$, showing consistency of the model.
Although there is no flavor violation in the muon sector, violation
of $e$ and $\tau$ lepton numbers do arise in the model.  Owing to
the violation of GIM mechanism in the left--handed lepton sector,
there is a $\overline{e} \tau Z$ coupling in the model. However,
this coupling is of order $(m_\mu m_\tau/m_E^2) \sim 10^{-10}$,
which is too small to be observed. In Section 5 we discuss lepton
number violating $\tau$ decays arising from $e-\tau-E$ mixing
and mediated by the $SU(2)_F$ gauge bosons.

Since $\tan\beta$ will turn out to be rather large in the model,
finite SUSY loop correction arising through chargino and neutralino
exchange are important for the $\tau$ lepton mass \cite{hall}. These corrections
are indicated in Eq. (\ref{ppp}) as $\Delta_\tau$. The tau mass
corrections are dominated by diagrams involving exchange of the
Bino/slepton and Higgsino/slepton. We use the following approximate
expressions for $\Delta_\tau$ in our numerical analysis
\cite{babu2}:
\begin{eqnarray}
\Delta_\tau&\simeq&\mu\tan\beta\left[\frac{\alpha_2}{8\pi}
M_2(I(\mu^2,\,m_{\tilde{\tau}}^2,\,M_2^2)+2I(\mu^2,\,m_{\tilde{\nu}}^2,\,M_2^2))+\frac{3\alpha_1}{20\pi}M_1
I(m_{\tilde{\tau}}^2,\,m_{\tilde{\tau}^c}^2,\,M_1^2)
\right.\nonumber\\
&-&\left.\frac{3\alpha_1}{40\pi}
M_1(I(\mu^2,\,m_{\tilde{\tau}}^2,\,M_1^2)-2I(\mu^2,\,m_{\tilde{\tau}^c}^2,\,M_1^2))\right].
\end{eqnarray}
Here the function $I$ is defined as
\begin{eqnarray}
I(m_1^2,\,m_2^2,\,m_3^2,\,)=\frac{1}{m_3^2}\left[\frac{x\,\ell n\,
x}{1-x}-\frac{y\,\ell n\, y}{1-y}\right]\frac{1}{x-y},
\end{eqnarray}
with $x=m_1^2/m_3^2$ and $y=m_2^2/m_3^2$.

The corrected running bottom quark mass is given by \cite{hall}
\begin{eqnarray}
m_b=y_b\upsilon_d (1+\Delta_b).
\end{eqnarray}
Here $\Delta_b$ is the finite SUSY loop correction arising from the
exchange of gluino and charginos and is given by
\begin{equation}
\Delta_b \simeq\mu\tan\beta\left[\frac{2\alpha_3}{3\pi}M_3
I(m_{\tilde{b}_1}^2,\,m_{\tilde{b}_2}^2,\,M_3^2) +
\frac{y_t^2}{16\pi^2}A_t
I(m_{\tilde{t}_1}^2,\,m_{\tilde{t}_2}^2,\,\mu^2)\right].
\end{equation}
\section{The SUSY spectrum\label{sec24}}
In this section we present the results for the masses of the SUSY
scalars. We will show  that the tachyonic slepton problem of AMSB is
cured by virtue of the positive contribution from the $SU(2)_{F}$
gauge sector to the masses for the first two families. For the
$\tau$ family sleptons, contributions arising from the Yukawa
coupling, which are significant for large $\tan\beta$, render the
squared masses positive.

\subsection{Slepton masses}
The slepton masses are obtained from a 2 $\times$ 2 mass matrix for
the ($\tilde{\mu},\,\tilde{\mu}^c$) sector (since the muon family
decouples from the rest of the sleptons) and from a 6 $\times$ 6
mass matrix for the \{$e,\tau,E, e^c, \tau^c, E^c$\} fields. For the
scalar muons the mass matrix is
\begin{eqnarray}\label{SU2349}
 M^2_{\tilde{\mu}} =\pmatrix{m^2_{\tilde{\mu}}
& m_{\mu}\left(A_{f_{e\mu}}-\mu\tan{\beta}\right)\cr
m_{\mu}\left(A_{f_{e\mu}}-\mu\tan{\beta}\right)&
m^2_{\tilde{\mu}^{c}}},
\end{eqnarray}
where the diagonal entries in the AMSB scheme are
\begin{eqnarray}\label{SU225}
m^2_{\tilde{\mu}}&=&\frac{M_{aux}^2}{(16\pi^2)}\left[2f_{e\mu}\beta(f_{e\mu})-\left(\frac{3}{2}g_{2}\beta(g_{2})
+\frac{3}{10}g_{1}\beta(g_{1})+\frac{3}{2}g_F\beta(g_F)\right)\right]+m_{\mu}^2+\frac{{g_F}^2}{4}(u_u^2-u_d^2),\nn\\
m^2_{\tilde{\mu}^{c}}&=&\frac{M_{aux}^2}{(16\pi^2)}\left[2f_{e\mu}\beta(f_{e\mu})-\left(\frac{6}{5}g_{1}\beta(g_{1})+\frac{3}{2}g_F\beta(g_F)\right)\right]+m_{\mu}^2+\frac{{g_F}^2}{4}(u_d^2-u_u^2).
\end{eqnarray}
 Note  the positive
contributions arising from the $SU(2)_{F}$ gauge sector given by the
term $-\frac{3}{2}g_F\beta(g_F)$, with gauge beta function
$\beta(g_F)=-\frac{3}{16\pi^2}g_F^3$. For ${g_F}\geqslant 0.9$, we
find the squared masses of all sleptons can be positive. It is
important to point out that the $SU(2)_F$ $D$--term contribution to
the diagonal entries of the mass matrix Eq. (\ref{SU2349}) is
positive for one slepton and negative for the other. Consistency
demands that this contribution be rather small compared to the other
terms. Thus $\tan\beta' \simeq 1$ is required in the model.

The mass matrix for the sleptons other than the scalar muons has the
form
\begin{eqnarray}\label{dc}
\mathcal{M}^2={\pmatrix{m^2_{\tilde{e}}&0&f_{e\mu}f_{eE}\upsilon_du_d&{\cal
A}&0 &0\cr 0&m^2_{\tilde{\tau}}&M_Ef_{\tau E}\upsilon_d&0&{\cal B}
&{\cal C}\cr f_{e\mu}f_{eE}\upsilon_du_d&M_Ef_{\tau
E}\upsilon_d&m^2_{\tilde{E}}&{\cal D}&0&M_EB_E\cr {\cal A} &0&{\cal
D} &m^2_{\tilde{e}^c}&0&M_Ef_{eE}u_d\cr 0 &{\cal
B}&0&0&m^2_{\tilde{\tau}^c}&f_{\tau}f_{\tau E}\upsilon_d^2\cr
0&{\cal C}&M_EB_E&M_Ef_{eE}u_d&f_{\tau}f_{\tau
E}\upsilon_d^2&m^2_{\tilde{E}^c}}},
\end{eqnarray}
where we have defined
\begin{eqnarray}\label{SU226}
m^2_{\tilde{e}}&=&\frac{M_{aux}^2}{(16\pi^2)}\left[2f_{e\mu}\beta(f_{e\mu})-\left(\frac{3}{2}g_{2}\beta(g_{2})
+\frac{3}{10}g_{1}\beta(g_{1})+\frac{3}{2}g_F\beta(g_F)\right)\right]\nn\\&+&f_{e\mu}^2\upsilon_d^2+\frac{{g_F}^2}{4}(u_d^2-u_u^2),\nn\\
m^2_{\tilde{e}^{c}}&=&\frac{M_{aux}^2}{(16\pi^2)}\left[2f_{e\mu}\beta(f_{e\mu})-\left(\frac{6}{5}g_{1}\beta(g_{1})+\frac{3}{2}g_F\beta(g_F)\right)\right]\nn\\
&+&f_{e\mu}^2\upsilon_d^2+f_{eE}^2u_u^2+\frac{{g_F}^2}{4}(u_u^2-u_d^2)\nn\\
m^2_{\tilde{\tau}}&=&\frac{M_{aux}^2}{(16\pi^2)}\left[f_{\tau}\beta(f_{\tau})+f_{\tau
E}\beta(f_{\tau
E})-\left(\frac{3}{10}g_{1}\beta(g_{1})+\frac{3}{2}g_{2}\beta(g_{2})\right)\right]+(f_{\tau}^2+f_{\tau E}^2)\upsilon_d^2,\nn\\
m^2_{\tilde{\tau}^{c}}&=&\frac{M_{aux}^2}{(16\pi^2)}\left[2f_{\tau}\beta(f_{\tau})-\left(\frac{6}{5}g_{1}\beta(g_{1})\right)\right]+f_{\tau}^2\upsilon_d^2\nn\\
m^2_{\tilde{E}}&=&\frac{M_{aux}^2}{(16\pi^2)}\left[f_{e E}\beta(f_{e
E})-\left(\frac{6}{5}g_{1}\beta(g_{1})\right)\right]+m_{E}^2+f_{eE}^2u_u^2,\nn\\
m^2_{\tilde{E}^{c}}&=&\frac{M_{aux}^2}{(16\pi^2)}\left[f_{\tau
e}\beta(f_{e\tau
})-\left(\frac{6}{5}g_{1}\beta(g_{1})\right)\right]+m_{E}^2+f_{\tau
E}^2\upsilon_d^2 \nonumber \\
{\cal A}&=&f_{e\mu}(A_{e\mu}\upsilon_d+\mu\upsilon_{u})\nn\\
{\cal B} &=&f_\tau (A_\tau\upsilon_d+\mu \upsilon_u)\nn\\
{\cal C} &=&f_{\tau E}(A_{\tau E}\upsilon_{d}+\mu \upsilon_u)\nn\\
{\cal D} &=&f_{eE}(A_{eE}u_d+\mu^{\prime}u_u)~.
\end{eqnarray}
The requirement that the slepton masses are positive puts
constraints on the couplings $f_\tau,\,\,f_{eE},\,\,f_{\tau e}$. We
find  $f_{\tau},\,f_{eE},\,f_{\tau E}\geq 0.5$ are needed.

The $\Psi_N$ scalar masses are give by
\begin{eqnarray}
m_{\tilde{N}_1}&=&\frac{M_{aux}^2}{(16\pi^2)}\left[-\frac{3}{2}g_F\beta(g_F)\right]+\frac{{g_F}^2}{4}(u_d^2-u_u^2)\nn\\
m_{\tilde{N}_2}&=&\frac{M_{aux}^2}{(16\pi^2)}\left[-\frac{3}{2}g_F\beta(g_F)\right]+\frac{{g_F}^2}{4}(u_u^2-u_d^2).
\end{eqnarray}
\subsection{Squark  masses}
The mixing matrix for the squark sector is identical to the usual
MSSM with no contributions from the $SU(2)_F$ sector. The diagonal
entries of the up and the down squark mass matrices are given by
\begin{eqnarray}\label{SU2H35}
m_{\tilde{U}_{i}}^2&=&(m^{2}_{soft})_{\tilde{Q}_{i}}^{\tilde{Q}_{i}}+m_{U_{i}}^{2}+\frac{1}{6}\left(4M_{W}^{2}-M_{Z}^{2}\right)\cos{2\beta},\nn\\
m_{\tilde{U}^{c}_{i}}^2&=&(m^{2}_{soft})_{\tilde{U}_{i}^{c}}^{\tilde{U}_{i}^{c}}+m_{U_{i}}^{2}-\frac{2}{3}\left(M_{W}^{2}-M_{Z}^{2}\right)\cos{2\beta},\nn\\
m_{\tilde{D}_{i}}^2&=&(m^{2}_{soft})_{\tilde{Q}_{i}}^{\tilde{Q}_{i}}+m_{D_{i}}^{2}-\frac{1}{6}\left(2M_{W}^{2}+M_{Z}^{2}\right)\cos{2\beta},\nn\\
m_{\tilde{D}_{i}^{c}}^2&=&(m^{2}_{soft})_{\tilde{D}_{i}^{c}}^{\tilde{D}_{i}^{c}}+
m_{D_{i}}^{2}+\frac{1}{3}\left(M_{W}^{2}-M_{Z}^{2}\right)\cos{2\beta},
\end{eqnarray}
were $m_{U_{i}}$ and $m_{D_{i}}$ are the quark masses of the
different generations with $i$ = 1, 2, 3. The soft masses are
obtained in AMSB from the RGE  as
\begin{eqnarray}\label{SU2H36}
(m^{2}_{soft})_{\tilde{Q}_{i}}^{\tilde{Q}_{i}}=\frac{M_{aux}^{2}}{16\pi^2}\left(Y_{u_{i}}
\beta{(Y_{u_{i}})}+Y_{d_{i}}\beta{(Y_{d_{i}})}-\frac{1}{30}g_{1}\beta{(g_{1})}-\frac{3}{2}g_{2}
\beta{(g_{2})} -\frac{8}{3}g_{3}\beta{(g_{3})}\right),
\end{eqnarray}
\begin{eqnarray}\label{SU2H37}
(m^{2}_{soft})_{\tilde{U}_{i}^{c}}^{\tilde{U}_{i}^{c}}&=&\frac{M_{aux}^{2}}{16\pi^2}\left(2Y_{u_{i}}
\beta{(Y_{u_{i}})}-\frac{8}{15}g_{1}\beta{(g_{1})}-\frac{8}{3}g_{3}\beta{(g_{3})}\right),\\
(m^{2}_{soft})_{\tilde{D}_{i}^{c}}^{\tilde{D}_{i}^{c}}&=&\frac{M_{aux}^{2}}{16\pi^2}\left(2Y_{d_{i}}
\beta{(Y_{d_{i}})}-\frac{2}{15}g_{1}\beta{(g_{1})}-\frac{8}{3}g_{3}\beta{(g_{3})}\right).
\end{eqnarray}
The RGE for quark sector Yukawa couplings are listed in Appendix A.2.

\section{Numerical results\label{sec25}}
Here we present our numerical results for the SUSY spectrum. Our
analysis follows the procedure of Ref. \cite{abg}.  The input values
of the SM gauge couplings \cite {Particle} used are:
\begin{eqnarray}
 &&\alpha^{-1}_{EM}(M_Z)= 128.91\pm 0.02,\nonumber \\
 &&\sin^2\theta_W(M_Z)=0.23120\pm 0.00015,\nonumber \\
 &&\alpha_3(M_Z)=0.1182\pm 0.0027.
\end{eqnarray}
We extrapolate these couplings to $M_{\rm SUSY} \simeq 1$ TeV using the
SM renormalization group equations.  We use the central value of the top mass taken to be $M_t=174.3$
GeV.

The scale of SUSY breaking, $M_{aux}$, should be in the range
$40-100$ TeV in order for the sparticle  masses to be in the range
$0.2-2$ TeV. Since the positivity of the mass--squared of the
slepton of the third family depends  on the Yukawa couplings, we
find that the couplings should obey $f_{\tau},\,f_{\tau E},~f_{eE}
\geq 0.5$.  This will lead to a large value of $\tan\beta \geq 40$.
For the positivity of the first two family slepton masses, the
$SU(2)_F$ gauge coupling should obey $g_F \geq 0.9.$

The parameters of the model are highly constrained.  Besides the
positivity of the slepton squared masses, one should ensure that the
lightest Higgs boson mass is above the current experimental limit,
$m_H \geq 114$ GeV. Furthermore, symmetry breaking should be
consistently achieved with the hierarchy $u_{u} \simeq u_{d} \gg
\upsilon_u \gg \upsilon_d$.  This hierarchy is needed to guarantee
that the $SU(2)_F$ gauge bosons are heavier than the $W$ and the $Z$
bosons.

We have not performed a systematic parameter search within the
model. By performing a ``spot search" we were able to find
consistent solutions. If we ``move around" a solution, we found that
the solution quickly disappears.  This feature indicates that the model
is highly predictive.

The first two family fermion masses do not play any significant role
in our fit, but the third family fermion masses do.  We choose
$m_\tau(m_\tau) = 1.777$ GeV as an input.  The running mass
$m_{\tau}(M_{\rm SUSY}) = 1.769$ GeV.  The input value of $b$-quark
mass is taken to be $m_b(m_b) = 4.8$ GeV, corresponding to
$m_b(M_{\rm SUSY}) = 2.8$ GeV.  In computing $\tau$ and $b$ masses
we include the finite SUSY loop corrections as noted in Eqs. (20)
and (23).

We present a specific fit in Table 2.  The parameters used for the
fit are indicated in the Table caption.  As shown in the Table, all
slepton squared masses are positive. An interesting feature of the
model is that the $\tilde{e}$ and $\tilde{\mu}$ are not nearly
degenerate.  For example, $m_{\tilde{e^c}} = 538$ GeV, while
$m_{\tilde{\mu^c}} = 834$ GeV.  Non-degenerate sleptons is a
characteristic feature of our model.

The lightest neutral Higgs boson mass is $m_H = 120$ GeV in our fit.
This is obtained after including the leading one--loop and two--loop
radiative corrections for $m_H$.  We follow the procedure outlined
in Ref. \cite{chaber}.  Since the entire SUSY spectrum is relatively
heavy, including the pseudoscalar Higgs boson, we decouple all SUSY
particles at $M_{\rm SUSY} = 1$ TeV, and use the SM RGE to compute
the evolution of the Higgs quartic coupling between 1 TeV and $m_t$.
There is a light scalar
$m_H' = 86 $ GeV, but this is mostly from the $\{\phi_u,~\phi_d\}$
sector and has very weak couplings to the SM fermions and gauge
bosons.

The lightest supersymmetric particle is found to be the neutral Wino
which is nearly mass degenerate with one of the charginos.  (The
mass splitting between the two is about 230 MeV \cite{Fengm}.) The
three $SU(2)_F$ gauge bosons have a common mass is found to be $\sim
1.91$ TeV with this set of input parameters. The heavy Higgs bosons,
Higgsinos and squarks masses are in the range $0.6-2.0$ TeV.

\begin{center}
\begin{table}[h!]
\begin{center}\small{
\begin{tabular}{|l|c|c|}\hline
\rule[1.5mm]{0mm}{0pt}Particles & Symbol& Mass (TeV)\\\hline
\rule[1.5mm]{0mm}{0pt}Neutralinos&$\{m_{\tilde{\chi}_{1}^{0}},\,\,m_{\tilde{\chi}_{2}^{0}},\,\,m_{\tilde{\chi}_{3}^{0}},\,\,m_{\tilde{\chi}_{4}^{0}}\}$&$\{0.149,\,\,0.235,\,\,0.614,\,\,0.912\}$\\\hline
\rule[1.5mm]{0mm}{0pt}Neutralinos&$\{m_{\tilde{\chi}_{5}^{0}},\,\,m_{\tilde{\chi}_{6}^{0}},\,\,m_{\tilde{\chi}_{7}^{0}}\}$&$\{0.917,\,\,1.593,\,\,2.452\}$\\\hline
\rule[1.5mm]{0mm}{0pt}Charginos&$\{m_{\tilde{\chi}_{1}^{\pm}},\,\,m_{\tilde{\chi}_{2}^{\pm}}\}$&$\{0.149,\,\,0.915\}$\\\hline
\rule[1.5mm]{0mm}{0pt}Charginos
($SU(2)_F$)&$\{m_{\tilde{\chi}_{1}^{\pm}},\,\,m_{\tilde{\chi}_{2}^{\pm}}\}$&$\{1.585,\,\,2.457\}$\\\hline
\rule[1.5mm]{0mm}{0pt}Gluino&$M_{3}$&$1.319$\\\hline
\rule[1.5mm]{0mm}{0pt}Neutral Higgs bosons
&$\{m_{h},\,\,m_{H},\,\,m_{A}\}$&$\{0.120,\,\,0.910,\,\,0.910\}$\\\hline
\rule[1.5mm]{0mm}{0pt}Neutral Higgs bosons
&$\{m_{h^\prime},\,\,m_{H^\prime},\,\,m_{A^\prime}\}$&$\{0.086,\,\,1.318,\,\,2.319\}$\\\hline
\rule[1.5mm]{0mm}{0pt}Charged Higgs bosons
&$m_{H^{\pm}}$&$0.913$\\\hline
 \rule[1.5mm]{0mm}{0pt}Charged Higgs
bosons $SU(2)_F$ &$m_{\phi^{\prime,\,\prime \star}}$&$2.321$\\\hline
 \rule[1.5mm]{0mm}{0pt}R.H smuon
&$\{m_{\tilde{\mu}_{1}}\}$&$\{0.834\}$\\\hline
\rule[1.5mm]{0mm}{0pt}L.H smuon
&$\{m_{\tilde{\mu}_{2}}\}$&$\{0.628\}$\\\hline
\rule[4.0mm]{0mm}{0pt}R.H sleptons
&$\{m_{\tilde{e}_{R}},\,\,m_{\tilde{\tau}_{1}},\,\,m_{\tilde{E}_{R}}\}$&$\{0.538,\,\,0.207,\,\,2.522\}$\\\hline
\rule[4.0mm]{0mm}{0pt}L.H sleptons
&$\{m_{\tilde{e}_{L}},\,\,m_{\tilde{\tau}_{2}},\,\,m_{\tilde{E}_{L}}\}$&$\{0.636,\,\,0.401,\,\,1.018\}$\\\hline
\rule[4.0mm]{0mm}{0pt}Sneutrinos&$\{m_{\tilde{\nu}_{e}},\,\,m_{\tilde{\nu}_{\mu}},\,\,m_{\tilde{\nu}_{\tau}}\}$&$\{0.636,\,\,0.834,\,\,0.414\}$\\\hline
\rule[4.0mm]{0mm}{0pt}Scalar $\Psi_N$
&$\{m_{\tilde{N}_1},\,\,m_{\tilde{N}_2}\}$&$\{0.673,\,\,0.863\}$\\\hline
\rule[1.5mm]{0mm}{0pt}R.H down squarks
&$\{m_{\tilde{d}_{R}},\,\,m_{\tilde{s}_{R}},\,\,m_{\tilde{b}_{1}}\}$&$\{1.241,\,\,1.241,\,\,1.171\}$\\\hline
\rule[1.5mm]{0mm}{0pt}L.H down squarks
&$\{m_{\tilde{d}_{L}},\,\,m_{\tilde{s}_{L}},\,\,m_{\tilde{b}_{2}}\}$&$\{1.231,\,\,1.231,\,\,1.016\}$\\\hline
\rule[1.5mm]{0mm}{0pt}R.H up squarks
&$\{m_{\tilde{u}_{R}},\,\,m_{\tilde{c}_{R}},\,\,m_{\tilde{t}_{1}}\}$&$\{1.233,\,\,1.233,\,\,0.946\}$\\\hline
\rule[1.5mm]{0mm}{0pt}L.H up squarks
&$\{m_{\tilde{u}_{L}},\,\,m_{\tilde{c}_{L}},\,\,m_{\tilde{t}_{2}}\}$&$\{1.228,\,\,1.228,\,\,1.133\}$\\
\hline \rule[4.0mm]{0mm}{0pt}$SU(2)_{F}$ gauge boson
&$M_Z^{\prime}$&$1.910$\\
\hline
\end{tabular}
\caption[Sparticle masses in the $SU(2)_F$ model for one choice of
input parameters.]%
{\footnotesize \small{Sparticle masses in Model 1 for the choice
$M_{aux}=57.605$ TeV, $y_{b}=0.95$, $f_{\tau}=0.55$, $f_{eE}=1.2$,
 $f_{\tau E}=0.53$, $g_F=1.0$, $M_{E}=0.0149$ TeV and $M_t=0.1743$ TeV,
 $u=2.702$ TeV, $\tan{\beta}=58.2$, $\tan{\beta^\prime}=1.08$,
$\mu =-0.908$ TeV, $\mu^\prime =0.236$ TeV, $B=0.016$ TeV,
$B^\prime=-3.676$ TeV, $B_E=0.007$ TeV.}}
  \label{SU2HTABLE2}}
  \end{center}
\end{table}
\end{center}

\section{Experimental implications\label{sec26}}
In this section we list the salient experimental signatures of the
model.

(i) Non-degeneracy of the first two family sleptons is a
characteristic feature of our model.  This is unlike most models of
supersymmetry breaking.  The origin of this splitting can be traced
back to the Yukawa couplings and the $SU(2)_F$ $D$-terms.

(ii) The model predicts large value of $\tan\beta \geq 40$.  There
are observable experimental consequences, which will be discussed
below.

(iii) Three degenerate vector gauge bosons with masses of order TeV
are predicted by the model.  These gauge bosons do not mix with the
$Z$ boson, nor do they couple to quarks.  Experimental discovery of
these bosons will be hard at a hadron collider, but should be easy
at a lepton collider.  Electroweak precision observables are left
intact by the new gauge sector, since there is no $Z-V$ mixing.  As
noted earlier, the presence of an approximate $Z_4$ symmetry in the
model prevents $\mu \rightarrow 3e$ and $\mu \rightarrow e \gamma$
decays that could have been mediated by the $V$ gauge bosons. For
the fit given in Table 2, the $SU(2)_F$ gauge bosons are degenerate
with a mass $M_V=1.910$ TeV. The most stringent
 constraint on $M_V$ arises from the process
 $e^+e^-\rightarrow\mu^+\mu^-$. LEP II has set severe
  constraints on lepton compositeness \cite{Eichten,Particle} from this
  process. For $\Lambda\,(ee\mu\mu) >9.5$ TeV
\cite{Eichten,Particle}, we obtain the limit $M_V>1.6$ TeV (for
$g_F = 1.0$). This limit is satisfied in our model.

(iv) Because $\tau$ and $e$ lepton numbers are not conserved in the
model, one would expect decays such as $\tau \rightarrow 3e$ and
$\tau \rightarrow e \mu^+ \mu^-$. These are mediated by the $V$
gauge bosons.  Note that the leptonic mass matrix of Eq. (18) has
both $e-E$ and $\tau-E$ mixings.  We denote by
$\theta_{e\tau}^{L,R}$ the $(e,\tau)$ entry of the matrix
$O_{L,R}^T.diag[1,-1,0,0].O_{L,R}$, where $O_L^T M_\ell O_R =
M_\ell^{\rm diag}$.  For our fit, these mixing angles are found to
be $\theta_{e\tau}^L = -2.1 \times 10^{-4}$ and $\theta_{e\tau}^R =
-2.3 \times 10^{-3}$. Since $\theta_{e\tau}^L$ is an order of
magnitude larger than $\theta_{e\tau}^R$, we ignore the latter and obtain the following
approximate expressions for the decay rates $\Gamma(\tau\rightarrow
3e)$ and $\Gamma(\tau\rightarrow \mu^+\mu^-e)$:
\begin{eqnarray}
\Gamma(\tau\rightarrow
3e)&\simeq&\frac{3}{8(192\pi^3)}\frac{g_F^4}{16}\frac{m_\tau^5}{M_V^4}|\theta^L_{e\tau}|^2,\\
\Gamma(\tau\rightarrow
\mu^+\mu^-e)&\simeq&\frac{1}{(192\pi^3)}\frac{g_F^4}{16}\frac{m_\tau^5}{M_V^4}|\theta^L_{e\tau}|^2.
\end{eqnarray}
We find the Branching ratios $Br\,(\tau\rightarrow 3e)\simeq
6.9\times 10^{-14}$ and $Br\,(\tau\rightarrow \mu^+\mu^-e)\simeq
1.84 \times 10^{-13}$.  These are clearly well below the current experimental sensitivity.

(v) The decay $\tau \rightarrow e \gamma$ is mediated by SUSY scalar
exchange. The dominant contribution to this decay amplitude arises
from the exchange of Bino and sleptons.  The rate for the decay is
given by
\begin{eqnarray}
\Gamma(\tau\rightarrow
e\gamma)&\simeq &\frac{\alpha_1}{4}m_\tau^3\left|\frac{3\alpha_1}{20\pi
m_{\tilde{\tau}}^2}m_{\tilde{B}}F(m_{\tilde{B}}^2/m_{\tilde{\tau}}^2)(\delta_{\tau
e}^{LR})\right|^2
\end{eqnarray}
where (in the standard notation) $\delta_{\tau e}^{LR}=\delta_{\tau
\tau}^{LR}\times \delta_{\tau e}^{RR}$ which is found to be 0.044
(0.719 $\times 0.061$) in our model and
\begin{eqnarray}
 F(x)&=&\frac{(1+4x-5x^2+4x\,\ell n(x)+2x^2\,\ell
n(x))}{2(1-x)^4}.
\end{eqnarray}
We find the branching ratio of $Br(\tau\rightarrow
e\gamma)=3.74\times 10^{-8}$ which is very close to the current
experimental limit and within reach of future experiments.

(vi) Since $\tan\beta$ is large, the Higgs boson mediated decay $B_s
\rightarrow \mu^+ \mu^-$ \cite{babu3} has a large rate.  To estimate
this we follow the analysis of Ref. \cite{babu3}.
\begin{eqnarray}
 BR(B^0\rightarrow \mu^+\mu^-)\simeq
\frac{\eta_{QCD}^2}{64\pi}\frac{m_B^3}{M_A^4}f_B^2\bar{y}_b^2y_{\mu}^2|V^*_{t(d,s)}V_{tb}|^2\chi_{FC}^2,
\end{eqnarray}
where $\chi_{FC}$ is given by
\begin{eqnarray}
\chi_{FC}=\frac{-\epsilon_u
y_t^2\tan{\beta}}{(1+\epsilon_g\tan{\beta})[1+(\epsilon_g+\epsilon_u
y_t^2)\tan{\beta}]},
\end{eqnarray}
$\epsilon_g$ and $\epsilon_u$ are given by
\begin{eqnarray}
\epsilon_u&=&\mu\tan\beta\left[\frac{2\alpha_3}{3\pi}M_3
I(m_{\tilde{b}_1}^2,\,m_{\tilde{b}_2}^2,\,M_3^2)\right], \nonumber\\
\epsilon_g &=&\mu\tan\beta\left[\frac{y_t^2}{16\pi^2}A_t
I(m_{\tilde{t}_1}^2,\,m_{\tilde{t}_2}^2,\,\mu^2)\right].
\end{eqnarray}
For the fit of Table 2, we find $\epsilon_g=-8.9 \times 10^{-3}$ and
$\epsilon_u=-2.1 \times 10^{-3}$ which leads to $\chi_{FC}=0.427$.
Consequently, the branching ratio $BR(B_s^0\rightarrow
\mu^+\mu^-)\simeq 7.1\times 10^{-8}$ (with $\eta_{QCD} = 1.5$), and using analogous expressions,
$BR(B_d^0\rightarrow \mu^+\mu^-)\simeq 2.8\times 10^{-9}$.  These
decays are within reach of ongoing experiments at the Tevatron
and/or the LHC.

(vii) The lightest R--odd SUSY particle in the model is the neutral
Wino ($\tilde{\chi}_1^0$) which is nearly mass degenerate with the
chargino. $\tilde{\chi}_1^0$  is stable and can be a candidate for
cold dark matter \cite{cold}.

 \section{Conclusion\label{sec27}}

In this paper we have presented a realistic supersymmetric model
based on a gauged $SU(2)$ family symmetry for the leptons.  The
$SU(2)$ symmetry is broken at the TeV scale along with
supersymmetry.  We have shown how such a scenario can be made
phenomenologically consistent.

In the context of anomaly mediated SUSY breaking, the model
presented provides a simple solution to the tachyonic slepton
problem.  Just as the color interactions make the squared masses of
squarks positive, the $SU(2)_F$ interactions make the slepton
squared masses positive.  A large value of $\tan\beta \geq 40$ is
predicted by the model, as needed for the positivity of the third
family slepton masses.

An intriguing feature of the model is that, although the first two
family sleptons are degenerate in mass in the  $SU(2)_F$ symmetric
limit, symmetry breaking effects render them non--degenerate.  This
is one of the few models where a non--degeneracy of first two family
sleptons is observed.  The SUSY spectrum is relatively heavy with
masses spanning the range 500 GeV - 2 TeV for most particles.  The
lightest $R$--odd particle is the neutral Wino, which is a candidate
for cold dark matter.

Other salient features of the model include observable rates for
$\tau \rightarrow e \gamma$ and $B_s \rightarrow \mu^+ \mu^-$
decays.

\appendix
\section{Appendix}

In this Appendix we give the one-loop anomalous dimensions for the
matter fields, beta-function for the gauge and Yukawa couplings and
for the soft SUSY breaking masses in the $SU(2)_F$ symmetric model.
\subsection{Anomalous
dimensions} The one--loop anomalous dimensions for the various
matters fields in our model are:
\begin{eqnarray}\label{SU2HAPPENDIX1}
16\pi^{2}\gamma_{\psi}&=&f_{e\mu}^2-\left(\frac{3}{10}g_{1}^{2}+\frac{3}{2}g_{2}^{2}+\frac{3}{2}g_F^{2}\right),\\
16\pi^{2}\gamma_{\psi^{c}}&=&2f_{e\mu}^2+f_{eE}^2-\left(\frac{6}{5}g_{1}^{2}+\frac{3}{2}g_F^{2}\right),\\
16\pi^{2}\gamma_{L_{\tau}}&=&f_{\tau}^2+f_{\tau
E}^2-\left(\frac{3}{10}g_{1}^{2}+\frac{3}{2}g_{2}^{2}\right),\\
16\pi^{2}\gamma_{\tau^{c}}&=&2f_{\tau}^2-\frac{6}{5}g_{1}^{2},\\
16\pi^{2}\gamma_{Q_{ij}}&=&(Y_dY_d^\dag)_{ji}+(Y_uY_u^\dag)_{ji}-\delta_i^j\left(\frac{1}{30}g_{1}^{2}+\frac{3}{2}g_{2}^{2}+\frac{8}{3}g_{3}^{2}\right),\\
16\pi^{2}\gamma_{U_{ij}}&=&2(Y_{u}^{\dag}Y_{u})_{ij}-\delta_{i}^{j}\left(\frac{8}{15}g_{1}^{2}+\frac{8}{3}g_{3}^{2}\right),\\
16\pi^{2}\gamma_{D_{ij}}&=&2(Y_{d}^{\dag}Y_{d})_{ij}-\delta_{i}^{j}\left(\frac{2}{15}g_{1}^{2}+\frac{8}{3}g_{3}^{2}\right),\\
16\pi^{2}\gamma_{H_{d}}&=&3Y_{b}^{2}+4f_{e\mu}^2+f_{\tau
E}^2+f_{\tau}^2-\frac{3}{10}g_{1}^{2}-\frac{3}{2}g_{2}^{2},\\
16\pi^{2}\gamma_{H_{u}}&=&3Y_{t}^{2}-\frac{3}{10}g_{1}^{2}-\frac{3}{2}g_{2}^{2},\\
16\pi^{2}\gamma_{\phi_d}&=&f_{e E}^2-\frac{3}{2}g_F^2,\\
16\pi^{2}\gamma_{\phi_u}&=&-\frac{3}{2}g_F^2,\\
16\pi^{2}\gamma_{E}&=&2f_{e E}^2-\frac{6}{5}g_{1}^2,\\
16\pi^{2}\gamma_{E^c}&=&2f_{\tau E}^2-\frac{6}{5}g_{1}^2.
\end{eqnarray}

\subsection{Beta functions} The beta functions for the Yukawa
couplings appearing in the superpotential, Eq. (4), are:
\begin{eqnarray}\label{SU2HAPPENDIX2}
\beta(Y_{b})&=&\frac{Y_{b}}{16\pi^2}\left(6Y_{b}^2+Y_{t}^2+f_{\tau}^2+f_{\tau
E}^2+4f_{e\mu}^2
-\frac{7}{15}g_{1}^{2}-3g_{2}^{2}-\frac{16}{3}g_{3}^{2}\right),\\
\beta(Y_{t})&=&\frac{Y_{t}}{16\pi^2}\left(6Y_{t}^2+Y_{b}^2
-\frac{13}{15}g_{1}^{2}-3g_{2}^{2}-\frac{16}{3}g_{3}^{2}\right),\\
\beta(Y_{\tau})&=&\frac{Y_{\tau}}{16\pi^2}\left(4Y_{\tau}^2+3Y_{b}^2+2f^2_{\tau
E}+2f_{e\mu}^2
-\frac{9}{5}g_{1}^{2}-3g_{2}^{2}\right),\\
\beta(f_{e E})&=&\frac{f_{e E}}{16\pi^2}\left(4f_{e
E}^2+2f_{e{\mu}}^2-\frac{12}{5}g_{1}^2
-3g_F^2\right),\\
\beta(f_{\tau E})&=&\frac{f_{\tau E}}{16\pi^2}\left(4f_{\tau
E}^2+2f_{\tau}^2+4f_{e\mu}^2+3Y_{b}^2-\frac{9}{5}g_{1}^2
-3g_{2}^2\right),\\
\beta(f_{e\mu})&=&\frac{f_{e\mu}}{16\pi^2}\left(7f_{e\mu}^2+2f_{\tau
E}^2+2f_{\tau}^2+2f_{e E}^2+3Y_{b}^2-\frac{9}{5}g_{1}^{2}-3g_{2}^{2}
-3g_F^2\right).
\end{eqnarray}

 The
gauge beta function of the model are
\begin{eqnarray}\label{SU24}
\beta(g_{i})&=&b_{i}\frac{g_{i}^{3}}{16\pi^2},
\end{eqnarray}
where $b_{i}=(\frac{39}{5}, 1, -3, -3)$ for $i=1-4$ with ${g_F}$
being the gauge coupling associated with the $SU(2)_F$ gauge group.

\subsection{$A$ terms} The trilinear soft SUSY breaking terms are
given by
\begin{eqnarray}\label{SU2HAPPENDIX4}
A_{Y}&=&-\frac{\beta{(Y)}}{Y}M_{aux},
\end{eqnarray}
where $Y=(Y_{u_i},\, Y_{d_i},\,Y_{l_{i}},\,f_{e E},\,f_{\tau
E},\,f_{\tau}$).

\subsection{Gaugino  masses}\label{appendix5} The
soft masses of the gauginos are given by:
\begin{eqnarray}\label{SU2HAPPENDIX5}
M_{i}&=&\frac{\beta{(g_{i})}}{g_{i}}M_{aux},
\end{eqnarray}
where $i=1,2,3,4$, corresponding to the gauge groups $U(1)_{Y}$,
$SU(2)_{L}$, $SU(3)_{C}$, $SU(2)_F$ with $\beta(g_i)$ given as in
Eq. (\ref{SU24}).

\subsection{Soft SUSY masses} The soft masses of the squarks and the
sleptons are given in the text. For the $H_u$, $H_d$, $\nu^c$,
$S_+$, $S_-$ fields they are:
\begin{eqnarray}\label{SU2HAPPENDIX6}
(\tilde{m}^{2}_{soft})_{H_{u}}^{H_{u}}&=&\frac{M_{aux}^{2}}{16\pi^2}\left(3Y_{t}
\beta{(Y_{t})}-\frac{3}{10}g_{1}\beta{(g_{1})}-\frac{3}{2}g_{2}\beta{(g_{2})}-2\left(\frac{x}{2}\right)^2g_F\beta{(g_F)}\right),\\
(\tilde{m}^{2}_{soft})_{H_{d}}^{H_{d}}&=&\frac{M_{aux}^{2}}{16\pi^2}\left(3Y_{b}
\beta{(Y_{b})}+Y_{\tau}\beta{(Y_{\tau})}+Y_{\tau E}\beta{(Y_{\tau
E})}-\frac{3}{10}g_{1}\beta{(g_{1})}-\frac{3}{2}g_{2}\beta{(g_{2})}\right.\nn\\
&-&\left. 2\left(-\frac{x}{2}\right)^2g_F\beta{(g_F)}\right),\\
(\tilde{m}^{2}_{soft})_{\phi_u}^{\phi_u}&=&\frac{M_{aux}^{2}}{16\pi^2}\left(-\frac{3}{2}g_F\beta{(g_F)}\right),\\
(\tilde{m}^{2}_{soft})_{\phi_d}^{\phi_d}&=&\frac{M_{aux}^{2}}{16\pi^2}\left(f_{e
E}\beta{(f_{e E})}-\frac{3}{2}g_F\beta{(g_F)}\right).
\end{eqnarray}
The soft mass parameters of the sleptons and the squarks are given
in the text.

\section*{Acknowledgments}
This work  is supported in part by DOE Grant \# DE-FG02-04ER-46140
and \# DE-FG02-04ER-41306 (OCA and KSB) and by \# DE-FG02-84ER40163
(IG).


\end{document}